\documentclass[12pt,a4paper]{article}
\usepackage{amsfonts,amssymb,amsmath}
\usepackage{color}
\usepackage{theorem,cite}
\newtheorem{definition}{Definition}[section]

\newtheorem{proposition}[definition]{Proposition}
\newtheorem{property}[definition]{Property}
\newtheorem{theorem}[definition]{Theorem}
\newtheorem{corollary}[definition]{Corollary}

\theorembodyfont{\rmfamily}
\newtheorem{rmk}{Remark}[section]

\numberwithin{equation}{section}

\definecolor{brique}{rgb}{.9,.2,0}
\definecolor{blvert}{rgb}{0,.8,.85}
\definecolor{vertcl}{rgb}{0,1,.7}
\newcommand\vertcl[1]{\textcolor{vertcl}{#1}}
\newcommand\blvert[1]{\textcolor{blvert}{#1}}
\newcommand\brique[1]{\textcolor{brique}{#1}}
\def\lapth{
\begin{picture}(142,70)(0,-15)\thicklines
\put(0,0){\vertcl{\rule{20pt}{4pt}}}
\put(19,1){\vertcl{\line(1,3){23}}} 
\put(20,1){\vertcl{\line(1,3){23}}} 
\put(21,1){\vertcl{\line(1,3){23}}}
\put(22,1){\vertcl{\line(1,3){23}}}
\put(45,70){\vertcl{\line(1,-3){23}}} 
\put(44,70){\vertcl{\line(1,-3){23}}} 
\put(43,70){\vertcl{\line(1,-3){23}}}
\put(42,70){\vertcl{\line(1,-3){23}}}
\put(2,24){\vertcl{\rule{120pt}{4pt}}}
\put(65,0){\vertcl{\rule{60pt}{4pt}}}
\put(5,37){\Huge{\brique{\textbf{L}}}} 
\put(62,37){\Huge{\brique{\textbf{PTh}}}}
\put(12,-8){\blvert{\rule{92pt}{3.5pt}}}
\put(24,-15){\blvert{\rule{57pt}{3.5pt}}}
\put(36,-22){\blvert{\rule{30pt}{3.5pt}}}
\end{picture}
\raisebox{35pt}{
\begin{minipage}{324pt}\begin{center}
\textbf{Laboratoire d'Annecy-le-Vieux de Physique
Th\'eorique}\\[4ex]
website: \texttt{http://lappweb.in2p3.fr/lapth-2005/}
\end{center}
\end{minipage}}\\
\vspace{10pt}\quad \hrulefill\\
\vspace{10pt}}

\newcommand{\nonu}{\nonumber \\}

\newcommand{\hs}[1]{\hspace{#1 mm}}

\newcommand{\eps}{\epsilon}

                    \def\cL{{\cal L}}
\def\cM{{\cal M}}          \def\cN{{\cal N}}          
                    \def\cR{{\cal R}}

\newcommand{\CC}{{\mathbb C}}

\newcommand{\II}{{\mathbb I}}

\newcommand{\MM}{\mbox{${\mathbb M}$}}

\newcommand{\ZZ}{{\mathbb Z}}

\newcommand{\prt}{\partial}

\newcommand{\wt}[1]{\widetilde{#1}}
\newcommand{\mb}[1]{\hs{4}\mbox{#1}\hs{4}}

\newcommand{\half}{\frac{1}{2}}
\newcommand{\sfrac}[2]{{\textstyle{\frac{#1}{#2}}}}

\newcommand{\prf}{\underline{Proof:}\ }
\newcommand{\finprf}{\null \hfill {\rule{5pt}{5pt}}\\[1.2ex]\indent}

\def\tr{\mathop{\rm tr}\nolimits}

\begin{document}

\markright{\today\dotfill DRAFT\dotfill }
\setcounter{page}{0}
\pagestyle{empty}
\newcommand{\LAP}{LAPTH}

\hspace{-1cm}\lapth

\vspace{20mm}

\begin{center}

%
{\LARGE{\sffamily Super-Hubbard models and applications}}\\[1cm]

{\large J.M. Drummond, G. Feverati, L. Frappat
and E. Ragoucy\footnote{drummond@lapp.in2p3.fr,
feverati@lapp.in2p3.fr, frappat@lapp.in2p3.fr, 
ragoucy@lapp.in2p3.fr}\\[.21cm] 
\textit{Laboratoire de Physique Th{\'e}orique \LAP\\[.242cm]
 LAPP, BP 110, F-74941  Annecy-le-Vieux Cedex, France. }}
\end{center}
\vfill\vfill

\begin{abstract}
We construct XX- and Hubbard- like models based on unitary superalgebras
$gl(N|M)$ generalising Shastry's and Maassarani's approach of the algebraic
case. We introduce the R-matrix of the $gl(N|M)$ XX model and that of
the Hubbard model defined by coupling two independent XX models. In both
cases, we show that the R-matrices satisfy the Yang--Baxter equation, we
derive the corresponding local Hamiltonian in the transfer matrix formalism
and we determine the symmetry of the Hamiltonian. Explicit examples are
worked out. In the cases of the $gl(1|2)$ and $gl(2|2)$ Hubbard models, a
perturbative calculation at two loops \emph{\`a la} Klein and Seitz is 
performed.
\end{abstract}

\vfill
\rightline{\tt hep-th/0703078}
\rightline{\LAP-1176/07}
\rightline{March 2007}

\newpage
 \pagestyle{plain}

\section{Introduction}

The Hubbard model was introduced in order to study strongly correlated
electrons \cite{Hubbard,Gutzwiller} and has been used to describe the Mott
metal-insulator transition \cite{Mott,Hubbard3}, high $T_c$
superconductivity \cite{Anderson,Affleck} and chemical properties of
aromatic molecules \cite{heilieb}. Since then, it has been widely studied,
essentially due to its connection with condensed matter physics. The
literature on the Hubbard model being rather large, we do not aim at being
exhaustive and rather refer to the books \cite{Monto,EFGKK} and references
therein. Exact results have been mostly obtained in the case of the
one-dimensional model, which enters the framework of our study. In
particular, the 1D model has been solved by means of the Bethe ansatz in
the celebrated paper by Lieb and Wu \cite{LiebWu}. However, the set of
eigenfunctions considered there was incomplete, and a complete set of
eigenstates was constructed in \cite{EKScomp} using the $SO(4)$ symmetry of
the 1D Hubbard Hamiltonian.

Although the Hubbard model certainly exhibits fascinating features among
integrable systems, the understanding of the model within the framework of
the quantum inverse scattering method appeared only in the mid eighties.
The R-matrix of the Hubbard model was first constructed by Shastry
\cite{shastry,JWshas} and Olmedilla et al. \cite{Akutsu}, by coupling
(decorated) R-matrices of two independent $XX$ models, through a term
depending on the coupling constant $U$ of the Hubbard potential. The proof
of the Yang--Baxter relation for the corresponding R-matrix was given by
Shiroishi and Wadati \cite{shiro2}. The construction of the
R-matrix was then generalised in the $gl(N)$ case by Maassarani et al.,
first for the XX model \cite{maasa} and then for the $gl(N)$ Hubbard model
\cite{maasa2,maasa3}. Within the QISM framework, the eigenvalues of the
transfer matrix of the Hubbard model were found using the algebraic
Bethe ansatz together with certain analytic properties in
\cite{YueDegu,RamMar,martins}.

One of the main motivations for the present study of the Hubbard model and
its generalisations is the fact that it has recently appeared in the
context of $N=4$ super Yang-Mills theory in two distinct ways. Firstly, it
was noticed in \cite{Rej:2005qt} that the Hubbard model at half-filling,
when treated perturbatively in the coupling, reproduces the long-ranged
integrable spin chain of \cite{Beisert:2004hm} as an effective theory. It
thus provides a localisation of the long-ranged spin chain model and
gives a potential solution to the problem of describing interactions
which are longer than the length of the spin chain. The Hamiltonian of this
chain was conjectured in \cite{Beisert:2004hm} to be an all-order
description of the dilatation operator of $N=4$ super Yang-Mills in the
$su(2)$ subsector. That is, the energies of the spin chain are conjectured
to be the anomalous dimensions of the gauge theory operators in this
subsector. In relation to this, an interesting approach to the Hubbard model is
given in \cite{ffgr} that leads to the evaluation of energies for the
antiferromagnetic state and allows one to control the order of the limits
of large coupling and large length of the operators/large angular momentum.

The Hubbard model has also arisen in a slightly different way in the
context of $N=4$ super Yang--Mills (SYM). Following reasoning developed in
\cite{staudacher04}, the long range spin chain describing $N=4$ super
Yang--Mills theory can be described in terms of scattering of
momentum-carrying excitations (at least in the limit of very long operators
or chains). Under the assumption of integrability, this scattering is
governed by a two particle scattering matrix which is essentially
determined up to an overall phase factor by $su(2|2)$ symmetry
\cite{Beisert:2005tm}. This phase factor was introduced in
\cite{Arutyunov:2004vx} where its importance for matching with data from the
string theory regime was discussed.
In a recent paper \cite{Beisert:2006qh} it has been
shown that the S-matrix thus derived satisfies the Yang--Baxter
relation (or a twisted version, see \cite{Arutyunov:2006yd})
and in fact is proportional to the tensor product of two copies of
Shastry's R-matrix \cite{shastry,JWshas}. The undetermined dressing phase
of the $S$-matrix can be constrained by appealing to crossing-symmetry
\cite{Janik}. A proposal for its complete form was given in \cite{BES}
based on an earlier guess \cite{Eden:2006rx} and conjectures for the form
of the string Bethe ansatz \cite{BHL}. The non-triviality of the dressing
phase leads to modifications of the proposal of \cite{Beisert:2004hm} at
four loops and beyond. Following the suggestion of \cite{BES} this leads to
transcendental contributions to the anomalous dimensions and thus
(presumably) to some modification of the underlying Hubbard model of 
\cite{Rej:2005qt}.

An interesting common feature of these observations is the relation of the
Hubbard model coupling to the Yang-Mills coupling. This raises the
possibility that there may be some integrable extension of the Hubbard
model which contains both elements as part of a larger description of $N=4$
super Yang-Mills theory. We will not construct such a model here but we
will discuss a general approach to constructing a number of supersymmetric
Hubbard models. Each of these models can be treated perturbatively and thus
gives rise to an integrable long-ranged spin chain as an effective theory.

Other supersymmetric generalisations of the Hubbard model have been
constructed, see e.g. \cite{EKS,BGLZ}. These approaches mainly concern high
$T_c$ superconductivity models and their relation with the $t-J$ model.
They essentially use the $gl(1|2)$ or $gl(2|2)$ superalgebras, which appear
as the symmetry algebras of the Hamiltonian of the model. Our approach
however is different and is based on the QISM framework. It ensures the
integrability of the model and allows one to obtain local Hubbard-like
Hamiltonians for general $gl(N|M)$ superalgebras. They can be interpreted in
terms of `electrons' after a Jordan--Wigner transformation.

The plan of the paper is as follows. In section \ref{sect:XX}, we define
supersymmetric XX models whose R-matrices are based on the unitary series
$gl(N|M)$. We introduce the corresponding Hamiltonians and determine the
symmetry of the model. In section \ref{sect:Hub}, we construct the
associated Hubbard-type model, mimicking the Shastry and Maassarani
construction. We prove the Yang--Baxter relation for the super Hubbard
R-matrix, which allows us to define the monodromy and transfer matrices.
The symmetry of the super Hubbard model based on $gl(N|M)$ is shown to be
$gl(N-1|M-1) \oplus gl(1|1) \oplus gl(N-1|M-1) \oplus gl(1|1)$. In section
\ref{sect:examples}, we give some examples, writing explicitly the
Hamiltonians in the $gl(2|2)$, $gl(1|2)$ and $gl(4|4)$ cases. In the first
two cases, we also perform a second order perturbation computation
\textit{\`a la} Klein and Seitz \cite{klstz} and note a relation with the
spectrum of the effective two-site Hamiltonian with the dilatation operator
in the $su(1|2)$ sector of $N=4$ SYM.

\section{Super XX models based on $gl(N|M)$\label{sect:XX}}
We follow the construction given in \cite{maasa,martins}, extending it
to the case of superalgebras. In the following, we note
$K=N+M$.

We will use the standard auxiliary space notation, i.e. to any matrix $A\in
End(\CC^K)$, we associate the matrices $A_{1}=A\otimes \II$ and
$A_{2}=\II\otimes A$ in $End(\CC^{K})\otimes End(\CC^K)$. More
generally, when considering equalities in $End(\CC^{K})^{\otimes k}$,
we take $A_{j}$, $j=1,\ldots,k$ to act trivially in all spaces
$End(\CC^K)$, but the $j^{th}$ one.

To deal with superalgebras, we will also need a $\ZZ_{2}$ grading $[.]$ on
indices $j$, such that $[j]=0$ will be associated to bosons and $[j]=1$ to
fermions. Accordingly, the elementary matrices $E_{ij}$ (with 1 at
position $(i,j)$ and 0 elsewhere) will have grade
$[E_{ij}]=[i]+[j]$, and we will use the super-trace:
\begin{equation}
\tr A = \sum_{j=1}^{K} (-1)^{[j]}\,A_{jj} \mb{if} A=\sum_{i,j=1}^{K}
A_{ij}\,E_{ij}\,.
\end{equation}
The grading we use is given by
\begin{equation}
{[j]}=\begin{cases} 0 \mb{for} 1\leq j\leq N, \\
1 \mb{for} N < j\leq N+M.
\end{cases}
\end{equation}

\subsection{R-matrix\label{sec:R-XX}}

The R-matrix of the $gl(N|M)$ XX model is defined as:
\begin{equation}
R_{12}(\lambda) = \Sigma_{12}\,P_{12} + \Sigma_{12}\,\sin\lambda +
(\II\otimes\II-\Sigma_{12})\,P_{12}\,\cos\lambda
\label{def:RXX}
\end{equation}
where $P_{12}$ is the permutation operator,
\begin{equation}
P_{12} = \sum_{i,j =1}^{ K} (-1)^{[j]}\, E_{ij} \otimes E_{ji}
\end{equation}
and $\Sigma_{12}$ is built from projection operators
\begin{eqnarray}
\Sigma_{12} &=& \pi_{1}\,\wt\pi_{2}+\wt\pi_{1}\,\pi_{2} \mb{with}
\pi=\sum_{j \ne N,K} E_{jj}\quad,\quad \wt\pi=\II-\pi = E_{NN}+E_{KK}
\,.\quad
\label{def:Sigma}
\end{eqnarray}
It is easy to show that $\Sigma_{12}$ is also a projector,
$\Sigma_{12}^2=\Sigma_{12}$.

Let us introduce the diagonal matrix $C$:
\begin{equation}
C = \sum_{j \ne N,K} E_{jj} - E_{NN} - E_{KK} =\pi-\wt\pi\,.
\label{eq:matC}
\end{equation}
This matrix obeys $C^{2}=\II$ and is related to the R-matrix through the equalities
\begin{equation}
\Sigma_{12}=\half(1-C_{1}C_{2}) \mb{and}
\II\otimes\II-\Sigma_{12}=\half(1+C_{1}C_{2})\,.
\end{equation}
One has
\begin{theorem}\label{theo:R-XX}
The matrix (\ref{def:RXX}) satisfies the following properties:
\begin{itemize}
\item[--]
$C$-invariance:
\begin{equation}
C_1 \, C_2 \, R_{12}(\lambda) = R_{12}(\lambda) \,  C_1 \, C_2
\label{eq:invar}
\end{equation}
\item[--]
$C$-parity:
\begin{equation}
R_{12}(-\lambda) = C_1 \, R_{12}(\lambda) \, C_2
\label{eq:antisym}
\end{equation}
\item[--]
Symmetry: 
\begin{equation}
R_{12}(\lambda) = R_{21}(\lambda) 
\label{eq:symm}
\end{equation}
\item[--]
Unitarity: 
\begin{equation}
R_{12}(\lambda) \, R_{21}(-\lambda) = (\cos^2 \lambda) \, \II \otimes
\II
\label{eq:unitarity}
\end{equation}
\item[--] Regularity : 
\begin{equation} R_{12}(0)=P_{12}\end{equation}
\item[--]
Exchange relation: 
\begin{equation}
R_{12}(\lambda) \, R_{21}(\mu) = R_{12}(\mu) \, R_{21}(\lambda)
\label{eq:exchange}
\end{equation}
\item[--]
Yang--Baxter equation (YBE): 
\begin{equation}
R_{12}(\lambda_{12})\,R_{13}(\lambda_{13})\,R_{23}(\lambda_{23}) = 
R_{23}(\lambda_{23})\,R_{13}(\lambda_{13})\,R_{12}(\lambda_{12})
\mb{where} \lambda_{ij} = \lambda_i-\lambda_j.
\label{eq:YBE}
\end{equation}
\item[--]
Decorated Yang--Baxter equation (dYBE): 
\begin{equation}
R_{12}(\lambda'_{12})\,C_1\,R_{13}(\lambda_{13})\,R_{23}(\lambda'_{23}) = 
R_{23}(\lambda'_{23})\,R_{13}(\lambda_{13})\,C_1\,R_{12}(\lambda'_{12})
\mb{with} \lambda'_{ij} = \lambda_i+\lambda_j.
\label{eq:decYBE}
\end{equation}
 \end{itemize}
\end{theorem}
\prf
$C$-invariance, $C$-parity, symmetry, unitarity relation, regularity 
and exchange relation follow from a direct calculation, using the properties
\begin{equation}
C^2=\II\quad;\quad C_1 \, \Sigma_{12} = \Sigma_{12}\, C_1
= -\Sigma_{12}\, C_2 =-C_2\, \Sigma_{12} \,.
\label{eq:pptC}
\end{equation}
 The decorated Yang--Baxter equation is a consequence of the Yang--Baxter 
equation and the invariance property. Indeed, the Yang--Baxter equation 
reads, with the change of variable $\lambda_2 \to -\lambda_2$,
\begin{equation}
R_{12}(\lambda'_{12})\,R_{13}(\lambda_{13})\,R_{23}(-\lambda'_{23}) = 
R_{23}(-\lambda'_{23})\,R_{13}(\lambda_{13})\,R_{12}(\lambda'_{12})
\,.
\end{equation}
Using the antisymmetry property (\ref{eq:antisym}), one gets
\begin{equation}
R_{12}(\lambda'_{12})\,R_{13}(\lambda_{13})\,C_2\,R_{23}(\lambda'_{23})\,C_3 = 
C_2\,R_{23}(\lambda'_{23})\,C_3\,R_{13}(\lambda_{13})\,R_{12}(\lambda'_{12})
\,.
\end{equation}
Multiplying this last equation by $C_1\,C_2$ on the left and by $C_3$ on
the right, and using the invariance property (\ref{eq:invar}), one obtains
(\ref{eq:decYBE}). It remains thus to show the YBE.\\
To prove YBE, one evaluates the difference LHS $-$ RHS of (\ref{eq:YBE}). 
One notes first that the
terms in $\Sigma_{ab}$, $\Sigma_{ab} P_{ab}$ and
$(\II\otimes\II-\Sigma_{ab})P_{ab}$ alone satisfy the Yang--Baxter
equation. The expression is further simplified using the fact that
$\Sigma_{ab}$ is a projector and ordering all terms with the $\Sigma$'s on
the left and the $P$'s on the right. One is left, after some algebra and
the use of standard trigonometric relations, with only two terms:
\begin{eqnarray}
&& \alpha_1 \big( (\Sigma_{12} \Sigma_{13} + \Sigma_{12} \Sigma_{23}
- \Sigma_{12}) P_{12} P_{13} - (\Sigma_{12} \Sigma_{23} + \Sigma_{13}
\Sigma_{23}  - \Sigma_{23}) P_{12} P_{23} \big) \\
&& \alpha_2 \big( \Sigma_{12} - \Sigma_{23} - \Sigma_{12} \Sigma_{13} +
\Sigma_{13} \Sigma_{23} \big) P_{13}
\end{eqnarray}
where $\alpha_1 = \sin\lambda_{12} \cos\lambda_{13} + \cos\lambda_{13}
\sin\lambda_{23} - \cos\lambda_{12} \sin\lambda_{13} \cos\lambda_{23}$ and
$\alpha_2 = \sin\lambda_{12} (1-\cos\lambda_{13}) \sin\lambda_{23}$. A
direct calculation of these two terms gives identically zero thanks to the
relation $\Sigma_{12}=\half(1-C_{1}C_{2})$. This ends the proof of YBE.
\finprf

\paragraph{Special case of $gl(1|1)$}\null\ \\
In the case of $gl(1|1)$ the above construction leads to a trivial 
R-matrix, because there is no index $j$ such that $j\neq N,K$. 
However, one can check that modifying the definitions of the 
projectors and $C$ according to
\begin{equation}
\pi=E_{11}\ ;\ \wt\pi=\II-\pi=E_{22}\ ;\ C=\pi-\wt\pi
\end{equation}
all the properties remain valid. The R-matrix keeps the same form 
(\ref{def:RXX}), with $\Sigma_{12}$ defined as in (\ref{def:Sigma}). 
We will use this R-matrix for this particular case. Explicitly, one has
\begin{eqnarray}
R(\lambda) &\!\!=\!\!& E_{21}\otimes E_{12}-E_{12}\otimes E_{21} +
\sin(\lambda)\,\left( E_{11}\otimes E_{22}+E_{22}\otimes E_{11}\right) 
\nonumber \\
&& + \cos(\lambda)\,\left( E_{11}\otimes E_{11}+E_{22}\otimes E_{22}
\right)\,.
\end{eqnarray}

\subsection{Monodromy and transfer matrices}
{F}rom the R-matrix, one constructs the ($L$ sites) monodromy matrix
\begin{equation}
\cL_{0<1\ldots L>}(\lambda) = R_{01}(\lambda)\,R_{02}(\lambda)\cdots
R_{0L}(\lambda)
\end{equation}
which obeys the relation
\begin{equation}
R_{00'}(\lambda-\mu)\, \cL_{0<1\ldots L>}(\lambda)\, \cL_{0'<1\ldots L>}(\mu) =
\cL_{0'<1\ldots L>}(\mu) \, \cL_{0<1\ldots L>}(\lambda)\,R_{00'}(\lambda-\mu)
\,.\ 
\label{RLL-XX}
\end{equation}
This relation allows us to construct an ($L$ sites) integrable XX spin
chain through the transfer matrix
\begin{equation}
t_{1\ldots L}(\lambda) = \tr_{0} \cL_{0<1\ldots L>}(\lambda) = \tr_{0}
\Big(R_{01}(\lambda)\,R_{02}(\lambda)\cdots R_{0L}(\lambda)\Big)\,,
\end{equation}
where $\tr_{0}$ is the super-trace in the auxiliary space 0. 
Indeed, the relation (\ref{RLL-XX}) implies that the transfer matrices for
different values of the spectral parameter commute
\begin{equation}
[t_{1\ldots L}(\lambda)\,,\,t_{1\ldots L}(\mu)]=0\,.
\end{equation}
Then, the XX-Hamiltonian is defined by
\begin{equation}
H=t_{1\ldots L}(0)^{-1}\, t'_{1\ldots L}(0)
\end{equation}
where the prime $'$ denotes the derivative w.r.t. $\lambda$. Since the
R-matrix is regular, $H$ is local:
\begin{equation}
H=\sum_{j=1}^{L} H_{j,j+1}\mb{with} H_{j,j+1}=P_{j,j+1}\,R'_{j,j+1}(0) =
P_{j,j+1}\,\Sigma_{j,j+1}
\label{eq:XXHam}
\end{equation}
where we have used periodic boundary conditions, i.e. identified the site
$L+1$ with the site 1. Explicitly, the two-site Hamiltonian reads
\begin{equation}
H_{j,j+1} = \sum_{i \ne N,K} \Big[ E_{iN} \otimes E_{Ni} - E_{iK} \otimes E_{Ki}
+ (-1)^{[i]} \big( E_{Ni} \otimes E_{iN} + E_{Ki} \otimes E_{iK} \big) 
\Big] \,. 
\end{equation}
 
\subsection{Symmetry of super XX models}
Starting from a general $K\times K$ matrix $\cM$ generating (a
representation of) the superalgebra $gl(N|M)$, a direct calculation shows
that for
\begin{equation}
\MM=\pi\,\cM\,\pi+\wt\pi\,\cM\,\wt\pi\in gl(N-1|M-1)\oplus gl(1|1)
\label{def:M}
\end{equation}
we have
\begin{equation}
(\MM_{1}+\MM_{2})\,R_{12}(\lambda) = R_{12}(\lambda)\,(\MM_{1}+\MM_{2})\,.
\end{equation}
In words, the R-matrix admits a $gl(N-1|M-1)\oplus gl(1|1)$ symmetry
superalgebra whose generators have the form
\begin{equation}
\begin{array}{l}
\displaystyle E_{jk}\ ,\ j,k\neq N,K \mb{for} gl(N-1|M-1)\\
\displaystyle E_{NN}\ ;\ E_{NK}\ ;\ E_{KN}\ ;\ E_{KK}\mb{for}
gl(1|1)\,.
\end{array}\label{eq:gen-glN-1}
\end{equation}
Let us remark \textit{en passant} that the associated symmetry group
is in fact a \textit{super}group, i.e. parameters entering the group
generators have to be graded according to the grading of the superalgebra.
This does not affect the `bosonic' subgroup $GL(N-1)\otimes GL(M-1)\otimes
U(1)\otimes U(1)$, but the (other) `fermionic' generators need to have
Grassmann valued parameters. Note that $C$-invariance is just a
particular case of the above (bosonic) symmetry group.
 
As a consequence, the transfer matrix also admits  $gl(N-1|M-1)\oplus gl(1|1)$
symmetry superalgebra, where the generators are given by 
\begin{equation}
\MM_{<1\ldots L>}=\MM_{1}+\MM_{2}+\ldots+\MM_{L},
\end{equation}
where $\MM$ is one of the generators given in (\ref{eq:gen-glN-1}).
The same is true for any Hamiltonian $H$ built on the transfer matrix.

The remaining generators which would allow one to enlarge the symmetry to 
a $gl(N|M)$ superalgebra are given by
\begin{equation}
V=\pi\,\cM\,\wt\pi+\wt\pi\,\cM\,\pi\mb{generated by} E_{jN}
\ ;\ E_{jK}\ ;\ E_{Nj}\ ;\ E_{Kj}\ , j\neq N,K\,.
\end{equation}
They obey
\begin{equation}
V\,C=-C\,V \mb{so that}
V_{1}\,\Sigma_{12}+\Sigma_{12}\,V_{1}=V_{1}\ ;\quad
V_{2}\,\Sigma_{12}+\Sigma_{12}\,V_{2}=V_{2}\,.
\end{equation}
This proves that
\begin{equation}
(V_{1}+V_{2})\,R_{12}(\lambda) = \wt R_{12}(\lambda)\,(V_{1}+V_{2})
\end{equation}
where $\wt R_{12}(\lambda)$ is deduced from $R_{12}(\lambda)$ by
exchanging $\Sigma_{12}$ and $\II\otimes\II-\Sigma_{12}$. Hence, $V$
is not associated to a symmetry of the R-matrix in the usual way. 

Note however that we have the relation
\begin{equation}
V_{1}\,V_{2}\,R_{12}(\lambda)=R_{12}(\lambda)\,V_{1}\,V_{2}\,.
\end{equation}
It induces a $gl(N|M)$ symmetry superalgebra for the XX Hamiltonian, with
generators $\MM_{1}\cdot \MM_{2}\cdots \MM_{L}$, where $\MM=E_{jk}$,
$1\leq j,k\leq K$. Unfortunately, the action of the generators on the
Hamiltonian eigenvectors is identically zero, except on the
pseudo-vacuum. This symmetry thus yields no information.

\subsection{Generalisations}
One can construct a more general R-matrix, defined by 
\begin{eqnarray}
R_{12}(\lambda;q_{1},q_{2},\eps_{1},\eps_{2}) &=& 
\hat\Sigma_{12}(q_{1},q_{2},\eps_{1},\eps_{2})\,\sin\lambda 
+ \Big(\Sigma_{12} + 
(\II\otimes\II-\Sigma_{12})\,\cos\lambda\Big)\,P_{12} \qquad
\label{def:R-XXgen}
\end{eqnarray}
where 
\begin{eqnarray}
 \hat\Sigma_{12}(q_{1},q_{2},\eps_{1},\eps_{2})
=\sum_{j<N}\Big\{ q_{1}E_{NN}\otimes E_{jj}+\frac{1}{q_{1}}
 E_{jj}\otimes E_{NN} +q_{2}E_{KK}\otimes E_{jj}+\frac{1}{q_{2}}
 E_{jj}\otimes E_{KK}\Big\}\nonu
+\sum_{N<j<K}\Big\{ 
\eps_{1}\Big(q_{1}E_{NN}\otimes E_{jj}+\frac{1}{q_{1}}
 E_{jj}\otimes E_{NN}\Big) +
 \eps_{2}\Big(q_{2}E_{KK}\otimes E_{jj}+\frac{1}{q_{2}}
 E_{jj}\otimes E_{KK}\Big)\Big\}\qquad
\end{eqnarray}
The parameters $q_{1},q_{2}$ are complex numbers, while 
$\eps_{1},\eps_{2}$ take values in $\{-1,1\}$.
One has
\begin{eqnarray*}
&&\hat\Sigma_{12}(1,1,1,1)=\Sigma_{12}\\
&&\hat \Sigma_{12}(q_{1},q_{2},\eps_{1},\eps_{2})\,
\hat \Sigma_{12}(p_{1},p_{2},\mu_{1},\mu_{2})
 = \hat\Sigma_{12}(q_{1}p_{1} ,q_{2}p_{2},\eps_{1}\mu_{1},\eps_{2}\mu_{2})\,.
\end{eqnarray*}
Note that only $\Sigma_{12}$ is a projector.

It can be checked that the theorem \ref{theo:R-XX} is also valid for the
R-matrix (\ref{def:R-XXgen}), except for the symmetry (\ref{eq:symm}) which
now reads
\begin{equation}
R_{21}(\lambda;q_{1},q_{2},\eps_{1},\eps_{2})
=R_{12}(\lambda;\frac{1}{q_{1}},\frac{1}{q_{2}},\eps_{1},\eps_{2}).
\end{equation}

In fact, $R_{12}(\lambda;q_{1},q_{2},\eps_{1},\eps_{2})$ is the (Drinfeld)
twist of $R_{12}(\lambda;1,1,1,\eps_{1}\eps_{2})$:
\begin{eqnarray}
&&D_{1}\,R_{12}(\lambda;q_{1},q_{2},\eps_{1},\eps_{2})\,D_{2}^{-1} =
R_{12}(\lambda;1,1,\mu\eps_{1},\mu\eps_{2})\nonu &&\mb{ with }
D=\frac{1}{q_{1}}E_{NN}+\frac{1}{q_{2}}E_{KK}+
\sum_{j<N}E_{jj}+\mu\sum_{N<j<K}E_{jj}\,.
\end{eqnarray}
Since $D$ belongs\footnote{Strictly speaking it is
$\left(\frac{\mu^{M-1}}{q_{1}q_{2}}\right)^\frac{1}{K}\,D$ which belongs to
this group.} to the group $SU(N)\otimes SU(M)$, the properties proved above
(in particular the Yang-Baxter equation) remain valid for the matrix
$R_{12}(\lambda;q_{1},q_{2},\eps_{1},\eps_{1})$. In the same way, it is
sufficient to work with the matrix $R_{12}(\lambda;1,1,1,-1)$ to get the
properties of the matrices
$R_{12}(\lambda;q_{1},q_{2},\eps_{1},-\eps_{1})$, so that there are
essentially two different solutions, corresponding to the cases
$\eps_{1}=\eps_{2}$ and $\eps_{1}=-\eps_{2}$, hence defining two classes of
super XX spin chains.
Below, we will focus on the R-matrix built on $\Sigma_{12}$.

\section{Super-Hubbard models based on $gl(N|M)$\label{sect:Hub}}

We use the R-matrices defined above to build generalisations of the
Hubbard model. The usual Hubbard model is obtained when we specialise 
to the case of $gl(1|1)$.
We will use the results given in \cite{EFGKK}, generalising them to 
the case of superalgebras.

\subsection{R-matrix for super Hubbard models} 
One introduces the R-matrix of the super Hubbard model as the coupling of two super XX 
models, according to
\begin{equation}
R_{<12><34>}(\lambda_{1},\lambda_{2}) =
R_{13}(\lambda_{12})R_{24}(\lambda_{12}) +
\frac{\sin(\lambda_{12})}{\sin(\lambda'_{12})} \,\tanh(h'_{12})\,
R_{13}(\lambda'_{12})C_{1}R_{24}(\lambda'_{12})C_{2}
\label{def:R-XXfus}
\end{equation}
where again $\lambda_{12}=\lambda_{1}-\lambda_{2}$ and 
$\lambda'_{12}=\lambda_{1}+\lambda_{2}$. 
The definition of the parameter
$h'_{12}=h(\lambda_{1})+h(\lambda_{2})$ is given below.
It is easy to show that this 
R-matrix is symmetric
\begin{equation}
R_{<12><34>}(\lambda_{1},\lambda_{2}) =
R_{<34><12>}(\lambda_{1},\lambda_{2}) \,,
\end{equation}
regular
\begin{equation} 
R_{<12><34>}(\lambda_{1},\lambda_{1}) = P_{<12><34>}=P_{13}\,P_{24}
\end{equation}
and obeys the unitarity relation
\begin{eqnarray}
&&\hspace{-2.1ex}R_{<12><34>}(\lambda_{1},\lambda_{2})\,
R_{<34><12>}(\lambda_{2},\lambda_{1}) = 
\left( \cos^{4}(\lambda_{12})
-\Big(\frac{\sin(\lambda_{12})}{\sin(\lambda'_{12})}\,\tanh(h'_{12})\Big)^{2}\right)\,
\II_{<12>}\otimes\II_{<34>}\nonu
&&\mb{where} \II_{<12>}=\II\otimes\II\,.
\label{unit:R-XXfus}
\end{eqnarray}
\begin{property}
When the function $h(\lambda)$ is given by $\sinh(2h)=U\, \sin(2\lambda)$
for some (free) parameter $U$, the R-matrix (\ref{def:R-XXfus}) obeys YBE:
\begin{eqnarray}
&&R_{<12><34>}(\lambda_{1},\lambda_{2})\, 
R_{<12><56>}(\lambda_{1},\lambda_{3})\, 
R_{<34><56>}(\lambda_{2},\lambda_{3}) \nonu
&=& 
R_{<34><56>}(\lambda_{2},\lambda_{3})\, 
R_{<12><56>}(\lambda_{1},\lambda_{3})\, 
R_{<12><34>}(\lambda_{1},\lambda_{2})\,.
\end{eqnarray}
In that case, the coefficient in (\ref{unit:R-XXfus}) can be rewritten as
\begin{equation}
\cos^{2}(\lambda_{12})\,\left( \cos^{2}(\lambda_{12})
-\Big(\frac{\tanh(h_{12})}{\cos(\lambda'_{12})}\,\Big)^{2}\right)
\end{equation}
where $h_{12}=h(\lambda_{1})-h(\lambda_{2})$.
\end{property}
\prf 
We use a generalisation to superalgebras of the proof by Shiroishi
\cite{shiro}, following the proof for algebras presented in \cite{EFGKK}.
The starting point is the use \cite{Kore} of the following tetrahedral
relation \cite{Zam-tetra}:
\begin{equation}
R_{12}^a\, R_{13}^b\, R_{23}^c = 
\sum_{d,e,f=0}^{1} S^{abc}_{def}\, R_{23}^f\, R_{13}^e\,R_{12}^d
\ ,\quad \forall\ a,b,c=0,1
\label{eq:tetra-R}
\end{equation}
where $R_{jk}^{0}=R_{jk}(\lambda_{j}-\lambda_{k})$,
$R_{jk}^{1}=R_{jk}(\lambda_{j}+\lambda_{k})\,C_{j}$ and the R-matrix is
given by (\ref{def:RXX}). The non-vanishing entries of the matrix $S$ are
given by
\begin{equation}
\begin{array}{lll}\displaystyle
S^{0,0,0}_{0,0,0} =1\ ;&&\\[1.2ex]
S^{1,1,0}_{1,1,0} =1\ ;&   S^{0,1,1}_{0,1,1} =1\ ;&  S^{1,0,1}_{1,0,1}=1
\\[1.2ex]
S^{1,0,0}_{0,0,1} =V(\lambda_1,\lambda_2,-\lambda_3)\ ;
&  S^{1,0,0}_{0,1,0} =W(\lambda_1,\lambda_2,-\lambda_3)\
;& S^{1,0,0}_{1,1,1} =U(\lambda_1,\lambda_2,-\lambda_3) \\[1.2ex]
S^{0,1,0}_{0,0,1} =U(\lambda_1,-\lambda_2,\lambda_3) \ ;
&  S^{0,1,0}_{1,0,0} =W(\lambda_1,-\lambda_2,\lambda_3)
\ ;&  S^{0,1,0}_{1,1,1} =V(\lambda_1,-\lambda_2,\lambda_3)\\[1.2ex]
S^{0,0,1}_{0,1,0} =U(-\lambda_1,\lambda_2,\lambda_3) \ ;
&  S^{0,0,1}_{1,0,0} =V(-\lambda_1,\lambda_2,\lambda_3)
\ ;&  S^{0,0,1}_{1,1,1} =W(-\lambda_1,\lambda_2,\lambda_3) \\[1.2ex]
S^{1,1,1}_{0,0,1} =W(\lambda_1,\lambda_2,\lambda_3)\ ;&  
S^{1,1,1}_{0,1,0} =V(\lambda_1,\lambda_2,\lambda_3)\ ;& 
S^{1,1,1}_{1,0,0} =U(\lambda_1,\lambda_2,\lambda_3)
\end{array}
\end{equation}
with 
\begin{eqnarray}
&&U(\lambda_1,\lambda_2,\lambda_3) =
-\frac{\cos(\lambda_{13}')\,\sin(\lambda_{23}')}
{\sin(\lambda_{13})\,\cos(\lambda_{23})}\ ;\  
V(\lambda_1,\lambda_2,\lambda_3) =
-\frac{\sin(\lambda_{12}')\,\sin(\lambda_{23}')}
{\sin(\lambda_{12})\,\sin(\lambda_{23})}\ ;\quad\nonu
&&W(\lambda_1,\lambda_2,\lambda_3) =
\frac{\sin(\lambda_{12}')\,\cos(\lambda_{13}')}
{\sin(\lambda_{12})\,\cos(\lambda_{13})}\
;\ \lambda_{jk}=\lambda_{j}-\lambda_{k}\ ;\ 
\lambda_{jk}'=\lambda_{j}+\lambda_{k}\,,\ j,k=1,2,3\,.
\qquad
\end{eqnarray}
One needs also the relations:
\begin{eqnarray}
R_{23}^{1}\,R_{13}^{1}\,R_{12}^{1} &=& 
-\frac{\sin(\lambda_{13}')\,\cos(\lambda_{23}')}
{\sin(\lambda_{13})\,\cos(\lambda_{23})}
\,R_{23}^{0}\,R_{13}^{0}\,R_{12}^{1}
 +\frac{\cos(\lambda_{12}')\,\cos(\lambda_{23}')}
{\cos(\lambda_{12})\,\cos(\lambda_{23})}
\,R_{23}^{0}\,R_{13}^{1}\,R_{12}^{0}\nonu
&&   +\frac{\cos(\lambda_{12}')\,\sin(\lambda_{13}')}
{\cos(\lambda_{12})\,\sin(\lambda_{13})}
\,R_{23}^{1}\,R_{13}^{0}\,R_{12}^{0};
\label{eq:linRXX1}\\
R_{23}^{0}\,R_{13}^{0}\,R_{12}^{0}
&=& -\frac{\sin(\lambda_{13})\,\cos(\lambda_{23})}
{\sin(\lambda_{13}')\,\cos(\lambda_{23}')}
\,R_{23}^{1}\,R_{13}^{1}\,R_{12}^{0}
+\frac{\cos(\lambda_{12})\,\cos(\lambda_{23})}
{\cos(\lambda_{12}')\,\cos(\lambda_{23}')}
\,R_{23}^{1}\,R_{13}^{0}\,R_{12}^{1}\nonu
&& +\frac{\cos(\lambda_{12})\,\sin(\lambda_{13})}
{\cos(\lambda_{12}')\,\sin(\lambda_{13}')}
\,R_{23}^{0}\,R_{13}^{1}\,R_{12}^{1}\,.
\label{eq:linRXX2}
\end{eqnarray}
It is easy to prove, using for instance a symbolic computer program
\cite{form}, that all these relations hold for the R-matrix
(\ref{def:RXX}), provided $C_{j}^{2}=1$ and $P_{12}$ is a (super)
permutation operator.

Then, the end of the proof is similar to the algebra case: a direct (but
lengthy) calculation shows that the matrix
\begin{equation}
R_{<12><34>}(\lambda_{1},\lambda_{2}) = 
R_{13}(\lambda_{12})\, R_{24}(\lambda_{12}) 
+\alpha(\lambda_{1},\lambda_{2})\,
R_{13}(\lambda_{12}')\, R_{24}(\lambda_{12}') \,
C_{1}\,C_{2}
\end{equation}
obeys YBE provided the matrix $R_{12}(\lambda)$ obeys theorem
\ref{theo:R-XX}, relations (\ref{eq:tetra-R}) and
(\ref{eq:linRXX1}-\ref{eq:linRXX2}), and
$\alpha(\lambda_{1},\lambda_{2})$ is given by
\begin{equation}
\alpha(\lambda_{1},\lambda_{2})=
\frac{\cos(\lambda_{1}-\lambda_{2})\,\sinh(h_{1}-h_{2})}
{\cos(\lambda_{1}+\lambda_{2})\,\cosh(h_{1}+h_{2})}\,,
\end{equation}
where $h_{j}=h(\lambda_{j})$, $j=1,2$ is defined by 
$\sinh(2h)=U\, \sin(2\lambda)$ ; see \cite{EFGKK} for more details.
\finprf

\subsection{Monodromy matrices, transfer matrices and Hamiltonians} 
We remind the reader of the usual proof of integrability for
models based on transfer matrices. 
Let $\cR_{ab}(\lambda_{1},\lambda_{2})$ be an R-matrix 
obeying YBE, and being regular ($\cR_{ab}(\lambda,\lambda)=P_{ab}$). 
$a$ and $b$ denote the `coupled'
spaces ($a,b=<12>,<34>$ in the above cases). From YBE, one deduces
that the monodromy matrix
\begin{equation}
\cL_{a<b_{1}\ldots b_{L}>}(\lambda_{1},\lambda_{2})=
\cR_{ab_{1}}(\lambda_{1},\lambda_{2})\ldots
\cR_{ab_{L}}(\lambda_{1},\lambda_{2})
\end{equation}
obeys 
\begin{equation}
\cR_{aa'}(\lambda_{1},\lambda_{2})\cL_{a}(\lambda_{1},\lambda_{3})
\cL_{a'}(\lambda_{2},\lambda_{3}) = \cL_{a'}(\lambda_{2},\lambda_{3}) 
\cL_{a}(\lambda_{1},\lambda_{3})  \cR_{aa'}(\lambda_{1},\lambda_{2})
\end{equation}
where the dependence in the quantum spaces $b_{1},\ldots,b_{L}$ has
been omitted in $\cL$.
This relation proves that one can define a transfer matrix
\begin{equation}
\wt t(\lambda_{1},\lambda_{3}) =
\tr_{a}\cL_{a}(\lambda_{1},\lambda_{3})
\end{equation}
which obeys 
\begin{equation}
[\wt t(\lambda_{1},\lambda_{3})\,,\, \wt t(\lambda_{2},\lambda_{3})]=0\,.
\end{equation}
{F}rom the transfer matrix, one then deduces that all the Hamiltonians
\begin{equation}
H(\mu)=\wt t(0,\mu)^{-1}\,\left.\frac{\prt}{\prt\lambda}\wt
t(\lambda,\mu)\right\vert_{\lambda=0}
\label{def:Hub-gen}
\end{equation}
define, for any $\lambda$, an integrable model, since we have
\begin{equation}
[H(\mu),\wt t(\lambda,\mu)]=0\ ,\quad \forall \lambda\,.
\end{equation}
However, demanding further that the Hamiltonian be local, one is led (using
the regularity property) to specify $\mu=0$. One then gets
\begin{eqnarray}
 [H, t(\lambda)] = 0\ ,\quad \forall \lambda\ , \mb{for}
H = H(0)=t(0)^{-1}\,t'(0) 
\mb{and}  t(\lambda)=\wt t(\lambda,0)\,. \nonumber
\end{eqnarray}
The transfer matrix $t(\lambda)$ is constructed from the `reduced'
monodromy matrix
\begin{equation}
L_{a<b_{1}\ldots b_{L}>}(\lambda)=
\cR_{ab_{1}}(\lambda,0)\ldots \cR_{ab_{L}}(\lambda,0)\,.
\end{equation}
This `reduced' monodromy matrix is just the one used to define the
Hubbard model; one can compute 
\begin{equation}
R_{<12><34>}(\lambda,0) = R_{13}(\lambda)\,R_{24}(\lambda) \Big(\II\otimes\II
+\tanh(h)\,C_{1}\,C_{2}\Big)\,.
\end{equation}
Hence, it is the locality requirement that imposes the form of the
monodromy matrix used for the Hubbard model. More general (a priori
non local) Hamiltonians
can be defined using the form (\ref{def:Hub-gen}). 

The matricial form of the Hubbard-type Hamiltonian reads
\begin{equation}
H = \sum_{j=1}^{L}H_{<2j-1,2j><2j+1,2j+2>}=
\sum_{j=1}^{L}\check{R}_{<2j-1,2j><2j+1,2j+2>}'(0)
\label{eq:HubHam}
\end{equation}
with $\check{R}_{ab} = P_{ab} R_{ab}$, $\displaystyle \check{R}'_{ab}(0)
= \frac{d}{d\lambda} \check{R}_{ab}(\lambda,0) \Big\vert_{\lambda=0}$ and
\begin{equation}
H_{<2j-1,2j><2j+1,2j+2>}
=\Sigma_{2j-1,2j+1}P_{2j-1,2j+1}+\Sigma_{2j,2j+2}P_{2j,2j+2}
+\frac{U}{2}\,\Big(C_{2j-1}C_{2j}+C_{2j+1}C_{2j+2}\Big)
\label{eq:twositesham}
\end{equation}
where we have used periodic boundary conditions, i.e. identified sites
$<2L+1,2L+2>$ with sites $<1,2>$.

\begin{rmk}
Let us remark that, due to the coupling of the two XX models, 
the total number of sites is $2L$, but the number of `coupled' sites
(which are the real physical ones) is $L$. We will thus refer to
the Hubbard Hamiltonian (\ref{eq:HubHam}) as an $L$--site Hamiltonian.
This is consistent with the notation used after Jordan--Wigner
transformation (see below).
\end{rmk}

\subsubsection{Gauged version of the super Hubbard model} 
In the literature \cite{shastry,maasa2,maasa3},  a gauged version of the
Hubbard R-matrix is used. It
is defined by
\begin{eqnarray}
&&R^{\textrm{g}}_{<12><34>}(\lambda_{1},\lambda_{2}) = 
e^{\half h_{1}\,C_{1}C_{2}}\,e^{\half h_{2}\,C_{3}C_{4}}\,
R_{<12><34>}(\lambda_{1},\lambda_{2}) 
\, e^{-\half h_{1}\,C_{1}C_{2}}\, e^{-\half h_{2}\,C_{3}C_{4}}\nonu
&&\mb{where} h_{j}=h(\lambda_{j})\ ,\quad j=1,2
\label{def:RHub}
\end{eqnarray}
By construction, 
$R^{\textrm{g}}_{<12><34>}(\lambda_{1},\lambda_{2})$ also obeys YBE,
and is unitary, symmetric and regular. 

Following the same steps as before, we introduce the `reduced'
R-matrix
\begin{equation}
R^{\text{g}}_{<12><34>}(\lambda,0) = \frac{1}{\cosh(h)}\,
I_{12}(h)\,R_{13}(\lambda)\,R_{24}(\lambda)\,I_{12}(h)\,,
\end{equation}
where
\begin{equation}
I_{12}(h) = \cosh(\frac{h}{2})\,\II\otimes\II
+\sinh(\frac{h}{2})\,C_{1}\,C_{2}\,.
\end{equation}
It leads to the same Hamiltonian (\ref{eq:HubHam})--(\ref{eq:twositesham}).
This gauged version was originally introduced to recover the exact
form of Shastry's R-matrix.

\subsection{Symmetries} 
We generalise to superalgebras the results obtained for $su(N)$ Hubbard
models (see for instance \cite{maasa2,EFGKK}). For completeness, we compare
them with the well-known symmetry of the usual Hubbard model
\cite{heilieb,shiro}.
\begin{proposition}
The transfer matrix of the Hubbard model admits a
$gl(N-1|M-1)\oplus gl(1|1)\oplus gl(N-1|M-1)\oplus gl(1|1)$ symmetry
algebra, each of the $gl(N-1|M-1)\oplus gl(1|1)$ corresponding to the
symmetry of one XX model.

As a consequence this symmetry is also valid for the 
 Hubbard Hamiltonian.
\end{proposition}
\prf To prove this symmetry, it is
sufficient to remark that
\begin{equation}
\MM\,C=C\,\MM
\end{equation}
where $\MM$ is given in (\ref{def:M}). Thus, one gets 
\begin{equation}
[R_{<12><34>}(\lambda,0)\,,\, \MM_{1}+\MM_{3}]=0
=[R_{<12><34>}(\lambda,0)\,,\,\MM_{2}+\MM_{4}]
\end{equation}
where $R_{<12><34>}(\lambda,0)$ is the R-matrix of the 
Hubbard model.

As far as Hamiltonians are concerned, the generators of the symmetry 
have the form
\begin{equation}
\MM_{\textrm{evn}}=\sum_{j=1}^{L}\MM_{2j} \mb{and} 
\MM_{\textrm{odd}} =\sum_{j=1}^{L}\MM_{2j-1}
\end{equation}
They generate a $gl(N-1|M-1)\oplus gl(1|1)\oplus gl(N-1|M-1)\oplus 
gl(1|1)$ superalgebra.
\finprf
It is well-known that the Hubbard model possesses a $gl(2)\oplus gl(2)$, 
and thus it is natural to look for a $gl(N|M)\oplus gl(N|M)$ 
symmetry algebra for the generalised Hubbard models. Unfortunately, 
it seems not to be present. 
To discuss this point, we now review how the $so(4)$ symmetry
algebra is obtained in the framework of the Hubbard model and point
out some properties which are valid only in this case.

\subsubsection{Enhancement of the symmetry for Hubbard
  model\label{sec:gl11Hub}} 

As has been shown (originally in \cite{heilieb}, see also
\cite{EFGKK}), the full symmetry of the periodic Hubbard model (for finite
$L$) can be obtained through a change of the $\ZZ_{2}$-grading. In the
present context, it amounts to consider the $gl(1|1)$
superalgebra\footnote{It corresponds to the $\wt R$ matrix in the notation
of \cite{EFGKK}.}:
\begin{proposition}
In the $gl(1|1)$ case, the Hubbard R-matrix obeys
\begin{eqnarray}
&& (V^\pm_{12}-V^\pm_{34})\,R_{<12><34>}(\lambda_{1},\lambda_{2})
 = - R_{<12><34>}(\lambda_{1},\lambda_{2})\, (V^\pm_{12}-V^\pm_{34}) 
 \label{symVgl11}\\
&& (W^\pm_{12}+W^\pm_{34})\,R_{<12><34>}(\lambda_{1},\lambda_{2})
=  R_{<12><34>}(\lambda_{1},\lambda_{2})\,(W^\pm_{12}+W^\pm_{34}) 
 \label{symWgl11}
\end{eqnarray}
where $V^\pm=\sigma_\pm\otimes\sigma_{\pm}$ and 
$W^\pm=\sigma_\pm\otimes\sigma_{\mp}$.

These relations are not valid any more for a general  
 $gl(N|M)$ superalgebra for generators $V=E_{Nj}\otimes E_{Nj}$, 
$E_{Kj}\otimes E_{Kj}$, $E_{jN}\otimes E_{jN}$ or 
$E_{jK}\otimes E_{jK}$ and $W=E_{Nj}\otimes E_{jN}$, 
$E_{Kj}\otimes E_{jK}$, $E_{jN}\otimes E_{Nj}$ or 
$E_{jK}\otimes E_{Kj}$.
\end{proposition}
\prf
Direct calculation. In particular, we checked that this relation does 
not hold for $gl(1|2)$.\finprf

Relations (\ref{symVgl11})--(\ref{symWgl11}) are then enough 
to deduce the following corollary, proved in \cite{EFGKK}:
\begin{corollary}
For $gl(1|1)$, the Hamiltonian
\begin{equation}
H = \sum_{j=1}^{L} \check{R}_{<2j-1,2j><2j+1,2j+2>}'(0)
\end{equation}
possesses a $gl(2)\oplus gl(1)\oplus gl(1)$ symmetry algebra when $L$ is 
odd; this symmetry extends to a $gl(2)\oplus gl(2)$ algebra when $L$ 
is even.

The generators of this symmetry have the form
\begin{eqnarray}
&&S^{(V)}_{^\pm}=\sum_{j=1}^{L} (-1)^j\,V^\pm_{2j-1,2j}
\mb{and} S^{(W)}_{^\pm}=\sum_{j=1}^{L} W^\pm_{2j-1,2j},\\
&& S_{z}^{(V)} = \sum_{j=1}^{2L} C_{j}
\mb{and} S_{z}^{(W)} = \sum_{j=1}^{L} (C_{2j-1}-C_{2j}).
\end{eqnarray}
\end{corollary}

\subsubsection{Comparison with the `$gl(2)$ Hubbard model'\label{sec:gl2Hub}}

We give here the counterpart of the section \ref{sec:gl11Hub} when 
dealing with the $gl(2)$ Hubbard model, constructed using the transfer 
matrix approach.
\begin{proposition}\label{symgl2}
In the $gl(2)$ case, the Hubbard R-matrix obeys
\begin{eqnarray}
&& (V^\pm_{12}\,C_{3}C_{4}-V^\pm_{34})\,R_{<12><34>}(\lambda_{1},\lambda_{2})
= - R_{<12><34>}(\lambda_{1},\lambda_{2})\, 
(V^\pm_{12}-C_{1}C_{2}\,V^\pm_{34}) \label{com:VR}\\
&& (W^\pm_{12}\,C_{3}C_{4}+W^\pm_{34})\,R_{<12><34>}(\lambda_{1},\lambda_{2})
=  R_{<12><34>}(\lambda_{1},\lambda_{2})\, 
(W^\pm_{12}+C_{1}C_{2}\,W^\pm_{34}) \label{com:WR}
\end{eqnarray}
where $V^\pm=\sigma_\pm\otimes\sigma_{\pm}$ and 
$W^\pm=\sigma_\pm\otimes\sigma_{\mp}$.

These relations are not valid any more for a general $gl(N)$ algebra
for generators $V=E_{Nj}\otimes E_{Nj}$ or $E_{jN}\otimes E_{jN}$ 
and $W=E_{Nj}\otimes E_{jN}$ or $E_{jN}\otimes E_{Nj}$.
\end{proposition}
\prf
A direct calculation shows that the relations
$$
\begin{array}{lll}
\wt\pi_{1}\wt\pi_{2}P_{13}V^\pm_{14} = \wt\pi_{3}\wt\pi_{4}P_{24}V^\pm_{14} 
\quad ;
&\pi_{1}\pi_{2}P_{13}V^\pm_{41} = \pi_{1}\pi_{2}P_{24}V^\pm_{23} 
\quad ;\
&\pi_{1}\wt\pi_{2}P_{23}V^\pm_{24} = \wt\pi_{3}\pi_{1}P_{14}V^\pm_{12}
\\[1.2ex]
\wt\pi_{1}\wt\pi_{2}P_{24}W^\pm_{32} = \wt\pi_{3}\wt\pi_{2}P_{13}W^\pm_{34} 
\ ;\
&\pi_{1}\pi_{2}P_{13}W^\pm_{14} = \pi_{1}\pi_{4}P_{24}W^\pm_{34} 
\ ;\ 
&\pi_{1}\wt\pi_{2}P_{13}W^\pm_{14} = \pi_{1}\wt\pi_{2}P_{24}W^\pm_{32}
\end{array}
$$
hold for $gl(2)$, but not for the other (super)algebras. Using these 
relations, it is then easy to deduce the relations 
(\ref{com:VR})--(\ref{com:WR}). We also checked by direct calculation
that the relations  (\ref{com:VR})--(\ref{com:WR}) do not hold  for
$gl(3)$.\finprf
\begin{corollary}
For $gl(2)$, the Hamiltonians
\begin{equation}
H(\lambda_{1},\lambda_{2}) = \sum_{j=1}^{L-1} 
\check{R}_{<2j-1,2j><2j+1,2j+2>}(\lambda_{1},\lambda_{2})
\end{equation}
have a $gl(2)\oplus gl(2)$ symmetry algebra. 

It implies the same symmetry for the non-periodic Hubbard Hamiltonian
\begin{equation}
H_{n.p.} = 
\sum_{j=1}^{L-1} \check{R'}_{<2j-1,2j><2j+1,2j+2>}(0)\,.
\end{equation}
\end{corollary}
\prf
Multiplying from the left by $P_{13}\,P_{24}$,
the relations (\ref{com:VR})--(\ref{com:WR}) can be recast as
\begin{eqnarray}
&&(V^\pm_{12}-V^\pm_{34}\,C_{1}C_{2})\,\check R_{<12><34>}(\lambda_{1},\lambda_{2})
 = \check R_{<12><34>}(\lambda_{1},\lambda_{2})\, 
 (V^\pm_{12}-C_{1}C_{2}\,V^\pm_{34}),\\
&&(W^\pm_{12}+W^\pm_{34}\,C_{1}C_{2})\,\check R_{<12><34>}(\lambda_{1},\lambda_{2})
=  \check R_{<12><34>}(\lambda_{1},\lambda_{2})\, 
(W^\pm_{12}+C_{1}C_{2}\,W^\pm_{34})\qquad.
\end{eqnarray}
It shows that the generators (again with $V^\pm=\sigma_\pm\otimes\sigma_{\pm}$ and 
$W^\pm=\sigma_\pm\otimes\sigma_{\mp}$)
\begin{eqnarray}
V^\pm_{q} &=& \sum_{j=1}^{L}(-1)^j\,(C_{1}\ldots
C_{2j-2})\,V^\pm_{2j-1,2j}\nonu
&=&V^\pm_{12}-C_{1}C_{2}\,V^\pm_{34}+C_{1}C_{2}C_{3}C_{4}\,V^\pm_{56}
 -C_{1}C_{2}C_{3}C_{4}C_{5}C_{6}\,V^\pm_{78}+\ldots\\
W^\pm_{q} &=& \sum_{j=1}^{L}(C_{1}\ldots C_{2j-2})\,W^\pm_{2j-1,2j}\nonu
&=&W^\pm_{12}+C_{1}C_{2}\,W^\pm_{34}+C_{1}C_{2}C_{3}C_{4}\,W^\pm_{56}
+C_{1}C_{2}C_{3}C_{4}C_{5}C_{6}\,W^\pm_{78}+\ldots
\end{eqnarray}
commute with the above Hamiltonian (with no restriction on the parity 
of $L$). It is trivial to check that they
form a $gl(2)\oplus gl(2)$ algebra, with Cartan generators
\begin{equation}
S_{z}^\pm=\sum_{j=1}^{L}(C_{2j-1}\pm C_{2j}).
\end{equation}
\pagebreak[2]
\finprf

The Hamiltonians $H(\lambda_{1},\lambda_{2})$ and $H_{n.p.}$ are not
periodic, since they do not contain the term
$\check{R}_{<2L-1,2L><1,2>}(\lambda_{1},\lambda_{2})$, which breaks the
symmetry. Hence, $H_{n.p.}$ does not correspond to the usual Hubbard model.
However, in the thermodynamical limit $L\to\infty$, the missing periodic
term is sent to infinity, and one recovers the symmetry of the usual
Hubbard model.

\subsubsection{Jordan--Wigner transformation and periodicity}
Anticipating the reminder of section \ref{sect:JW} on Jordan--Wigner
transformation \cite{JW}, one is tempted to associate the $gl(2)$
construction of \cite{shastry,Akutsu} to the Hubbard model, but it is
well-known that, for algebras, the Jordan--Wigner transformation does not
preserve the periodic boundary condition \cite{JWshas} (see also
\cite{EFGKK}). Indeed, through this transformation, one gets for instance
\begin{equation}
c^\dag_{j}\,c_{j+1}\to E^{(j)}_{12}\, E_{21}^{(j+1)}\,,\ j=1,2,\ldots
\end{equation}
where the superscript indicates the site to which the matrices belong, and
the arrow denotes the Jordan--Wigner transformation. From periodicity, one
should thus get
\begin{equation}
c^\dag_{L}\,c_{1}\to E^{(L)}_{12}\, E_{21}^{(1)}
\end{equation}
However, performing the Jordan--Wigner transformation, one gets
\begin{equation}
c^\dag_{L}\,c_{1}\to E^{(L)}_{12}\, E_{21}^{(1)}\,(C_{1}\cdots C_{L-1})
\end{equation}
Hence, in the $gl(2)$ case, the Hubbard Hamiltonian we obtain is
non-periodic in terms of $c$ and $c^\dag$ (i.e. after
 Jordan--Wigner transformation).

When dealing with superalgebras, the Jordan--Wigner transformation is
modified \cite{EFGKK} (see a reminder in section \ref{sect:JW}), and now
respects the periodic boundary condition. Due to this,  we obtain the usual
(periodic) Hubbard Hamiltonian in the case of $gl(1|1)$.

In other words, for algebras, the
Jordan--Wigner transformation needs to modify the bosonic/fermionic
character of some operators: this is done using (non-local) products of $C_{j}\equiv
(1-2n_{j})$ generators which break the periodicity. For superalgebras, no
change of character is needed; the transformation is a local
isomorphism, so that periodicity is preserved.
In this respect, the superalgebra case looks more natural than the
algebraic one.

These considerations are consistent with the results of sections 
\ref{sec:gl11Hub} and \ref{sec:gl2Hub} about the 
symmetry of non-periodic $gl(2)$ and periodic $gl(1|1)$ Hubbard models.


\subsection{Change of notation}

The above presentation of the Hubbard model is based on the transfer matrix
formalism, the Hubbard model itself being obtained by coupling two
independent XX models, hence the notation used for the Hubbard Hamiltonian
(\ref{eq:HubHam}). In the following, we are dealing with explicit
expressions of this Hamiltonian in specific cases and we would like to make
contact with the notation commonly used in particular in the condensed
matter community. Therefore, we will perform a change of notation in the
rest of the paper in order to stick to more familiar expressions.

The construction of the Hubbard Hamiltonian, see eqs.
(\ref{eq:HubHam})--(\ref{eq:twositesham}), shows that one considers a $2L$
site lattice on which live two independent XX models, the first one living
on the odd sites, the second one on the even sites. We introduce a map on
the site labels in such a way that the $2L$ site lattice of the coupled XX
models is interpreted as a $L$ site lattice for the Hubbard model:
\begin{equation}
<2j-1,2j> \quad \to \quad j\uparrow \otimes\, j\downarrow \qquad 
(j=1,\ldots,L)
\label{eq:sitemap}
\end{equation}
the operators living on the odd (even) sublattice being labelled by
$\uparrow$ ($\downarrow$). With this notation, the Hubbard Hamiltonian 
(\ref{eq:HubHam})--(\ref{eq:twositesham}) reads 
\begin{equation}
H = \sum_{j=1}^{L} H_{j,j+1} \;\; \mbox{with} \;\; H_{j,j+1} =
\Sigma_{\uparrow,j,j+1} P_{\uparrow,j,j+1} + \Sigma_{\downarrow,j,j+1}
P_{\downarrow,j,j+1} + \frac{U}{2} \Big( C_{\uparrow,j} C_{\downarrow,j} +
C_{\uparrow,j+1} C_{\downarrow,j+1} \Big)
\end{equation}
where we used the periodicity conditions.

\subsection{Jordan--Wigner transformation \label{sect:JW}}

Let us consider $p$ sets of fermionic oscillators $c_{i}^{(q)},
c_{i}^{(q)\dagger}$ ($i=1,\ldots,L$ and $q=1,\ldots,p$) that satisfy the
usual anticommutation relations
\begin{equation}
\{ c_{i}^{(q)},c^{(q')\dagger}_{j} \} = \delta_{ij} \, \delta_{qq'} \qquad \{
c_{i}^{(q)},c_{j}^{(q')} \} = \{ c^{(q)\dagger}_{i},c^{(q')\dagger}_{j} \} =
0
\end{equation}
One defines the following matrix (where $n_{i}^{(q)} = c_{i}^{(q)\dagger}
c_{i}^{(q)}$ is the usual number operator)
\begin{equation}
X^{(q)}_i = \left( 
\begin{array}{cc}
1-n_{i}^{(q)} & c_{i}^{(q)} \\
c_{i}^{(q)\dagger} & n_{i}^{(q)} \\
\end{array}
\right)\,.
\end{equation}
The entries $X_{i;\alpha\beta}^{(q)}$ of this matrix have a natural grading
given by $[\alpha]+[\beta]$ where $[1]= 1$ and $[2] = 0$.

In the $gl(2^{p-1}|2^{p-1})$ case, one defines at each site $i$ the
generators
\begin{equation}
X_{i;\alpha_1\ldots\alpha_p,\alpha'_1\ldots\alpha'_p} = (-1)^{s} \,
X_{i;\alpha_1\alpha'_1}^{(1)} \; \ldots \; X_{i;\alpha_p\alpha'_p}^{(p)}
\;\; \mbox{where} \;\;
s = \sum_{a=2}^{p} [\alpha_a] \Big( \sum_{b=1}^{a-1} \big( [\alpha_b] +
[\alpha'_b] \big) \Big)\,.
\end{equation}
It is easy to verify the following properties:
\begin{eqnarray}
&& \big( X_{i;\alpha_1\ldots\alpha_p,\alpha'_1\ldots\alpha'_p} \big)^\dagger =
X_{i;\alpha'_1\ldots\alpha'_p,\alpha_1\ldots\alpha_p} \\
&& X_{i;\alpha_1\ldots\alpha_p,\alpha'_1\ldots\alpha'_p} \;
X_{i;\beta_1\ldots\beta_p,\beta'_1\ldots\beta'_p} = 
\delta_{\alpha'_1\beta_1} \ldots \delta_{\alpha'_p\beta_p} \,
X_{i;\alpha_1\ldots\alpha_p,\beta'_1\ldots\beta'_p} \\
&& \sum_{\alpha_1,\ldots,\alpha_p}
X_{i;\alpha_1\ldots\alpha_p,\alpha_1\ldots\alpha_p} = 1 \\
&& X_{i;\alpha_1\ldots\alpha_p,\alpha'_1\ldots\alpha'_p} \;
X_{j;\beta_1\ldots\beta_p,\beta'_1\ldots\beta'_p} = (-1)^{g}
X_{j;\beta_1\ldots\beta_p,\beta'_1\ldots\beta'_p} \;
X_{i;\alpha_1\ldots\alpha_p,\alpha'_1\ldots\alpha'_p} \qquad (i \ne j) \qquad \\
&& \mbox{where} \;\; g = \Big(\sum_{a=1}^p \big([\alpha_a] + [\alpha'_a]\big)\Big)
\Big(\sum_{b=1}^p \big([\beta_b] + [\beta'_b]\big)\Big)\,. \nonumber
\end{eqnarray}

The first three properties are local (on site) while the last one relates
different sites. They state that the
$X_{i;\alpha'_1,\ldots,\alpha'_p,\alpha_1,\ldots,\alpha_p}$ form an algebra
isomorphic to the tensor product of $L$ copies of $gl(2^{p-1}|2^{p-1})$.
The mapping $X_{\ldots}\rightarrow E_{\ldots} $ is known as a Jordan--Wigner
transformation. We observe that it can be uniquely defined once an entire
line $E^{(i)}_{1\alpha}$ is given; indeed, hermitian conjugation fixes the
corresponding column so the full matrix can be reconstructed in the
following steps:
\begin{enumerate}
\item
the element $E_{11}^{(i)}$ is associated to one of the $2^p$ diagonal 
generators $X_{i;\alpha_1,\ldots,\alpha_p,\alpha_1,\ldots,\alpha_p}$;
\item
the remaining $2^{p-1}-1$ bosonic generators are freely associated to the
bosonic ones $E^{(i)}_{1\alpha}$, $\alpha=2,\dots,2^{p-1}$;
\item 
the $2^{p-1}$ fermionic generators are freely associated to the 
$E^{(i)}_{1\alpha}$, $\alpha=2^{p-1}+1,\dots,2^p$.
\end{enumerate}
All specific realisations are isomorphic because they can be obtained one
from the other by exchanging lines and columns of the matrices. An example
of such a mapping is given in (\ref{eq:exempleJW}). The $gl(N|M)$ cases
that are not of the form $gl(2^{p-1}|2^{p-1})$ are ``incomplete'' and can
be obtained by embedding in the smallest algebra $gl(2^{p-1}|2^{p-1})$ such
that $N \le 2^{p-1} $ and $M \le 2^{p-1} $. Then, by removing an
appropriate choice of lines and columns, one projects the matrix
$X_{i;\alpha'_1,\ldots,\alpha'_p,\alpha_1,\ldots,\alpha_p}$ to a $gl(N|M)$
subalgebra. This can be done in many inequivalent ways and has been done in
section \ref{sect:gl12} with the projector (\ref{eq:projgl12}). In this
sense, any $gl(N|M)$ Hamiltonian describes a sector contained in the larger
$gl(2^{p-1}|2^{p-1})$ Hamiltonian's space of states.

\section{Examples\label{sect:examples}} 

\subsection{$gl(2|2)$ Hamiltonians}

In the $gl(2|2)$ case, the generators $X_{i;\alpha\beta,\alpha'\beta'}$
at each site $i$ are given by
\begin{equation}
X_{i;\alpha\beta,\alpha'\beta'} = (-1)^{([\alpha]+[\alpha'])[\beta]} \,
X_{i;\alpha\alpha'} \; X'_{i;\beta\beta'}\,.
\end{equation}
They are mapped on the $E_{pq}$ matrices with the following assignment of
indices ($\alpha,\beta,\alpha',\beta'=1,2$ and $p,q=1 \ldots 4$):
\begin{equation}
11 \to 1, \; 12 \to 3, \; 21 \to 4, \; 22 \to 2
\label{eq:exempleJW}
\end{equation}
which respects the grading in the sense that if
$(\alpha\beta,\alpha'\beta') \to (p,q)$, the grades of
$X_{\alpha\beta,\alpha'\beta'}$ and of $E_{pq}$ coincide. \\
Then the $gl(2|2)$ XX Hamiltonian (\ref{eq:XXHam}) reads as (the subscripts 
correspond to the site indices):
\begin{eqnarray}
H_{XX}^{gl(2|2)} &\!\!=\!\!& \sum_{i=1}^{L} \big( c_{i}^\dagger c_{i+1} +
c_{i+1}^\dagger c_{i} \big) \big( c_{i}'^\dagger c_{i+1}' +
c_{i+1}'^\dagger c_{i}' + 1 - n_{i}' - n_{i+1}' \big) \\
&=& \sum_{i=1}^{L} \left\{ - \, c_{i}'^\dagger c_{i}^\dagger c_{i+1}' c_{i+1} -
c_{i+1}'^\dagger c_{i+1}^\dagger c_{i}' c_{i} + c_{i}^\dagger
c_{i+1}'^\dagger c_{i}' c_{i+1} + c_{i}'^\dagger c_{i+1}^\dagger c_{i}
c_{i+1}' \right. \nonumber \\
&& \left. + \, \big( 1 - n_{i}' - n_{i+1}' \big) \big( c_{i}^\dagger
c_{i+1} + c_{i+1}^\dagger c_{i} \big) \right\}\,.
\label{eq:hamXX2}
\end{eqnarray}
This Hamiltonian exhibits interesting features. First of all, the number of
pairs (i.e. doubly occupied sites with one unprimed and one primed
particle) is conserved by the Hamiltonian, so that one can restrict the
study to sectors with a given number of pairs. The first two terms of
(\ref{eq:hamXX2}) correspond to a BCS-like conductivity in the physical
space (pair hopping), while the last term corresponds to ordinary
conductivity (hopping for unprimed particles with interaction with a
background of primed particles). The middle term corresponds to an exchange
between the two types of particles.

\medskip

As explained in section \ref{sect:Hub}, the Hubbard-type Hamiltonian
(\ref{eq:HubHam}) is obtained by coupling two copies of XX Hamiltonians, with
fermionic oscillators $c_{\sigma,i}^\dagger$ and $c_{\sigma,i}$,
$\sigma=\uparrow,\downarrow$. Hence, one gets for the $gl(2|2)$ Hubbard
Hamiltonian:
\begin{eqnarray}
H_{Hub}^{gl(2|2)} &\!\!=\!\!& \sum_{i=1}^{L} \; \Big\{ \;
\sum_{\sigma=\uparrow,\downarrow} \big( c_{\sigma,i}^\dagger c_{\sigma,i+1}
+ c_{\sigma,i+1}^\dagger c_{\sigma,i} \big) \big( c_{\sigma,i}'^\dagger
c_{\sigma,i+1}' + c_{\sigma,i+1}'^\dagger c_{\sigma,i}' + 1 - n_{\sigma,i}'
- n_{\sigma,i+1}' \big) \nonumber \\
&& + \, U (1-2n_{\uparrow,i})(1-2n_{\downarrow,i}) \; \Big\}\,.
\label{eq:hamhub22}
\end{eqnarray}
The space of states at each site $i$ is spanned by the vacuum
$|0\rangle_i$, the up states $|\!\uparrow\rangle_i$, $|\!\uparrow'\rangle_i$,
$|\!\uparrow\uparrow'\rangle_i$, the down states $|\!\downarrow\rangle_i$,
$|\!\downarrow'\rangle_i$, $|\!\downarrow\downarrow'\rangle_i$, and by tensoring
the up states with the down states, where $|\sigma\rangle_i \equiv
c_{\sigma,i}^\dagger |0\rangle_i$, $|\sigma'\rangle_i \equiv
c_{\sigma,i}'^\dagger |0\rangle_i$ and $|\sigma\sigma'\rangle_i \equiv
c_{\sigma,i}^\dagger c_{\sigma,i}'^\dagger |0\rangle_i$ with
$\sigma=\uparrow,\downarrow$. \\
This Hamiltonian can be compared with the $gl(4)$ Hubbard Hamiltonian 
which is given by
\begin{eqnarray}
H_{Hub}^{gl(4)} &\!\!=\!\!& \sum_{i=1}^{L} \; \Big\{ \;
\sum_{\sigma=\uparrow,\downarrow} \big( 
c_{\sigma,i}^\dagger c_{\sigma,i+1} c_{\sigma,i}'^\dagger c_{\sigma,i+1}'
+ c_{\sigma,i+1}^\dagger c_{\sigma,i} c_{\sigma,i+1}'^\dagger c_{\sigma,i}'
\nonumber \\
&& + \, n_{\sigma,i}' n_{\sigma,i+1}' (c_{\sigma,i}^\dagger
c_{\sigma,i+1}+c_{\sigma,i+1}^\dagger c_{\sigma,i}) + n_{\sigma,i}
n_{\sigma,i+1} (c_{\sigma,i}'^\dagger
c_{\sigma,i+1}'+c_{\sigma,i+1}'^\dagger c_{\sigma,i}') \big) \nonumber \\
&& + \, U
(1-2n_{\uparrow,i}n_{\uparrow,i}')(1-2n_{\downarrow,i}n_{\downarrow,i}') \;
\Big\}\,,
\end{eqnarray}
which is free of exchange terms.

\medskip

It is of interest to make a perturbative calculation of the $gl(2|2)$ Hubbard 
Hamiltonian (\ref{eq:hamhub22}) \emph{\`a la} Klein
and Seitz \cite{klstz}. To this aim, one introduces the notation
\begin{equation}
X_{ij} = \sum_{\sigma=\uparrow,\downarrow} c_{\sigma,i}^\dagger
c_{\sigma,j} \, \cN_{\sigma,ij}'
\label{eq:Xij}
\end{equation}
where $\cN_{\sigma,ij}' = c_{\sigma,i}'^\dagger c_{\sigma,j}' +
c_{\sigma,j}'^\dagger c_{\sigma,i}' + 1 - n_{\sigma,i}' - n_{\sigma,j}'$. 
The Hamiltonian takes then the form
\begin{equation}
H_{Hub}^{gl(2|2)} = \sum_{i=1}^{L} (X_{i,i+1}+X_{i+1,i}) + U \sum_{i=1}^{L} 
(1-2n_{\uparrow,i})(1-2n_{\downarrow,i})
\end{equation}
At large $U$, the potential term is the dominant one, while the $X$ term
can be treated as a perturbation. From the form of the potential term, one
is led to define a projector $\Pi_0$ on singly occupied states with unprimed
particles (i.e. $|\!\uparrow\rangle$ or $|\!\downarrow\rangle$), without any
limitations on the primed particles:
\begin{equation}
\Pi_0 = \prod_{i=1}^{L} (n_{\uparrow,i}-n_{\downarrow,i})^2\,.
\label{eq:projP0}
\end{equation}
Then, one can easily check that $X_{ij}^\dagger = X_{ji}$ and that $\Pi_0$
fulfills the following conditions:
\begin{equation}
\Pi_0 X_{ij} \Pi_0 = 0 \,, \qquad (1-\Pi_0) X_{ij} X_{ji} \Pi_0 = 0 \,, \qquad \Pi_0
X_{i,i+1} X_{i+1,i+2} \Pi_0 = 0 \,.
\label{eq:condP0}
\end{equation}
Note that the `dressing' factors $\cN_{\sigma,ij}'$ play no role in the
derivation of these relations since $\Pi_0$ and $\cN_{\sigma,ij}'$ commute.
It follows that the effective Hamiltonian at second order of perturbation
(the first order is vanishing) is given by
\begin{equation}
H_{\textrm{eff}}^{(2)} = - \frac{1}{2U}\sum_{i=1}^{L} \Pi_0 (X_{i,i+1}X_{i+1,i} +
X_{i+1,i}X_{i,i+1}) \Pi_0
\end{equation}
After some simple algebra, one finally gets
\begin{equation}
H_{\textrm{eff}}^{(2)} = - \frac{1}{U}\sum_{i=1}^{L} \Pi_0 \left[ (\half - 2
S_{i}^z S_{i+1}^z) - (S_{i}^+ S_{i+1}^- + S_{i}^- S_{i+1}^+) \,
\cN_{\uparrow,i i+1}' \, \cN_{\downarrow,i i+1}' \right] \Pi_0
\label{eq:hamhubeff}
\end{equation}
where one has defined the $sl(2)$ generators $S_{i}^+ =
c_{\uparrow,i}^\dagger c_{\downarrow,i}^{}$, $S_{i}^- =
c_{\downarrow,i}^\dagger c_{\uparrow,i}^{}$ and $S_{i}^z = \half
(n_{\uparrow,i} - n_{\downarrow,i})$.

\medskip

It is worthwhile to emphasise that the symmetry algebra of the Hubbard
Hamiltonian (\ref{eq:hamhub22}) is $gl(1|1) \oplus gl(1|1) \oplus gl(1|1)
\oplus gl(1|1)$, the symmetry algebra of the effective Hamiltonian at
second order of perturbation (\ref{eq:hamhubeff}) however is enhanced to
$gl(2|2) \oplus gl(2|2)$. Indeed, the following generators
\begin{eqnarray}
&& 
\rho_{\sigma}^+ = \sum_{i=1}^{L} n_{\sigma,i}^{} \, c_{\uparrow,i}'^\dagger \, c_{\downarrow,i}'^{} \;\;,\;\; 
\rho_{\sigma}^- = \sum_{i=1}^{L} n_{\sigma,i}^{} \, c_{\downarrow,i}'^\dagger \, c_{\uparrow,i}'^{} \;\;,\;\; 
\rho_{\sigma}^z = \sfrac12 \sum_{i=1}^{L} n_{\sigma,i}^{} (n_{\uparrow,i}'^{}-n_{\downarrow,i}'^{}) \;\;,\;\; 
N_{\sigma} = \sum_{i=1}^{L} n_{\sigma,i} 
\nonumber \\
&& 
\eta_{\sigma}^+ = \sum_{i=1}^{L} n_{\sigma,i}^{} \, c_{\uparrow,i}'^\dagger \, c_{\downarrow,i}'^\dagger \;\;,\;\; 
\eta_{\sigma}^- = \sum_{i=1}^{L} n_{\sigma,i}^{} \, c_{\downarrow,i}' \, c_{\uparrow,i}' \;\;,\;\; 
\eta_{\sigma}^z = \sfrac12 \sum_{i=1}^{L} n_{\sigma,i}^{} (n_{\uparrow,i}'^{}+n_{\downarrow,i}'^{}-1) 
\nonumber \\
&& 
\phi_{\sigma\tau}^+ = \sum_{i=1}^{L} n_{\sigma,i}^{} \, n_{-\tau,i}'^{} \, c_{\tau,i}'^\dagger \;\;,\;\; 
\phi_{\sigma\tau}^- = \sum_{i=1}^{L} n_{\sigma,i}^{} \, n_{-\tau,i}'^{} \, c_{\tau,i}' \qquad (\tau = 
\uparrow, \downarrow) 
\nonumber \\
&&
\chi_{\sigma\tau}^+ = \sum_{i=1}^{L} n_{\sigma,i}^{} (1-n_{-\tau,i}'^{}) \, c_{\tau,i}'^\dagger \;\;,\;\; 
\chi_{\sigma\tau}^- = \sum_{i=1}^{L} n_{\sigma,i}^{} (1-n_{-\tau,i}'^{}) \, c_{\tau,i}' \qquad (\tau = 
\uparrow,\downarrow) 
\label{eq:genegl22}
\end{eqnarray}
which generate two commuting copies of $sl(2|2)$ (one for
$\sigma=\uparrow$ and one 
for $\sigma=\downarrow$), commute with the effective Hamiltonian
(\ref{eq:hamhubeff}). Due to the presence of the projector $\Pi_0$ in
$H_{\textrm{eff}}^{(2)}$, the number operator $n_{\sigma,i}$ can be
expressed as $\half \pm S_i^z$ ($+$ for $\uparrow$ and $-$ for
$\downarrow$). The generators $\rho_\sigma^\pm, \rho_\sigma^z$ and
$\eta_\sigma^\pm, \eta_\sigma^z$ generate the four commuting $sl(2)$
algebras, and the remaining non-vanishing commutation relations are given
by
\begin{align}
& [ \rho_\sigma^\varepsilon,\phi_{\sigma\tau}^{\pm} ] = \pm
\phi_{\sigma,-\tau}^{\pm} \, \delta_{\varepsilon,\mp\tau} && [
\rho_\sigma^\varepsilon,\chi_{\sigma\tau}^{\pm} ] = \pm
\chi_{\sigma,-\tau}^{\pm} \, \delta_{\varepsilon,\mp\tau} \\
& [ \rho_\sigma^z,\phi_{\sigma\tau}^{\pm} ] = \pm\sfrac12\tau
\phi_{\sigma\tau}^{\pm} && [ \rho_\sigma^z,\chi_{\sigma\tau}^{\pm} ] =
\pm\sfrac12\tau \chi_{\sigma\tau}^{\pm} \\
& [ \eta_\sigma^\mp,\phi_{\sigma\tau}^{\pm} ] = \pm\tau
\chi_{\sigma,-\tau}^{\mp} && [ \eta_\sigma^\mp,\phi_{\sigma\tau}^{\pm} ] =
\pm\tau \chi_{\sigma,-\tau}^{\mp} \\
& [ \eta_\sigma^z,\phi_{\sigma\tau}^{\pm} ] = \pm\sfrac12
\phi_{\sigma\tau}^{\pm} && [ \eta_\sigma^z,\chi_{\sigma\tau}^{\pm} ] =
\pm\sfrac12 \chi_{\sigma\tau}^{\pm} \\
& \{ \chi_{\sigma\uparrow}^{\pm},\chi_{\sigma\downarrow}^{\mp} \} = - \{
\phi_{\sigma\uparrow}^{\pm},\phi_{\sigma\downarrow}^{\mp} \} = \rho_\sigma^\pm
&& \{ \phi_{\sigma\uparrow}^{\pm},\chi_{\sigma\downarrow}^{\pm} \} = - \{
\chi_{\sigma\uparrow}^{\pm},\phi_{\sigma\downarrow}^{\pm} \} =
\eta_\sigma^\pm \\
& \{ \phi_{\sigma\tau}^{+},\phi_{\sigma\tau}^{-} \} = \sfrac12 N_\sigma +
\eta_\sigma^z - \tau \rho_\sigma^z && \{
\chi_{\sigma\tau}^{+},\chi_{\sigma\tau}^{-} \} = \sfrac12 N_\sigma -
\eta_\sigma^z + \tau \rho_\sigma^z
\end{align}
where $\sigma,\tau = \uparrow, \downarrow$ and $\varepsilon = \pm$. \\
One has to add two more generators (the $U(1)$ factors such that one gets
two $gl(2|2)$ superalgebras), which are both represented in the present
realisation by matrices proportional to the unit matrix.

\subsubsection{Spectrum of the Hamiltonian and comparison with
$\mathcal{N}=4$-SYM}
It is interesting to check if (\ref{eq:hamhubeff}) has some relation with 
the dilatation operator of some sector in $\mathcal{N}=4$-SYM.
The proper candidate is  $su(1|2)$ sector whose two-site Hamiltonian is
$H_\text{SYM}=1-P_{12}$ \cite{beisert}.
On each site, after the projection $\Pi_0$, the two-site Hamiltonian 
(\ref{eq:hamhubeff}) acts
on eight states 
\begin{eqnarray}
&& |\!\uparrow\rangle,~  |\!\uparrow\uparrow'\rangle,~
|\!\uparrow\downarrow'\rangle,~ |\!\uparrow\uparrow'\downarrow'\rangle
\\[1.2ex]
&& |\!\downarrow\rangle,~  |\!\downarrow\uparrow'\rangle,~
|\!\downarrow\downarrow'\rangle,~ |\!\downarrow\uparrow'\downarrow'\rangle 
\end{eqnarray}
so that it is a $64\times 64$ matrix.
It has 32 vanishing lines and columns and the remaining part is built
of two-by-two blocks of  the form
\begin{eqnarray}\label{block}
&& B_{-}=\begin{pmatrix} 1& -1\\-1 &1
\end{pmatrix}  
\mb{or}
B_{+}=\begin{pmatrix} 1 &  1\\1 &1
\end{pmatrix}\,.
\end{eqnarray}
This means that the eigenvalue zero is represented 48 times and the
eigenvalue 2 appears 16 times.

The $H_\text{SYM}$ Hamiltonian is a $9\times 9$ matrix with two empty
lines and columns, three blocks $B_{-}$ and a one-dimensional diagonal
entry with value 2. Therefore, our Hamiltonian (\ref{eq:hamhubeff})
contains the correct $su(1|2)$ spectrum. The interpretation of states
is not obvious because the one-dimensional block with value 2 is
absent and we can obtain it only after a diagonalisation of one of the
blocks (\ref{block}) namely by mixing the states on two
sites. Moreover, the enhancement of symmetry seems to be strictly a
feature of this second order Hamiltonian and is most probably lost at
higher orders.

\subsection{$gl(1|2)$ Hamiltonians\label{sect:gl12}}

Following formula (\ref{eq:XXHam}), the $gl(1|2)$ XX-Hamiltonian can be
obtained from the $gl(2|2)$ one by suppressing the index 1 for example (or
equivalently the index 3). One gets therefore
\begin{equation}
H_{XX}^{gl(1|2)} = \sum_{i=1}^{L} \left\{ c_{i}'^\dagger c_{i+1}^\dagger
c_{i} c_{i+1}' + c_{i}^\dagger c_{i+1}'^\dagger c_{i}' c_{i+1} \, - \big(
c_{i}^\dagger c_{i+1} + c_{i+1}^\dagger c_{i} \big) n_{i}' n_{i+1}'
\right\}
\label{eq:hamXXgl12}
\end{equation}
where at each site $i$ the space of states is spanned by $c_{i}^\dagger
|0\rangle$, $c_{i}'^\dagger |0\rangle$ and $c_{i}^\dagger c_{i}'^\dagger
|0\rangle$. In other words, the space of states of the $gl(1|2)$ XX model
can be obtained from the space of states of the $gl(2|2)$ XX model by 
acting with the projector 
\begin{equation}
\Pi_{gl(1|2)} = \prod_{i=1}^{L} (n_{i}+n_{i}'-n_{i}n_{i}')
\label{eq:projgl12}
\end{equation}
It can be easily verified that one has
\begin{equation}
H_{XX}^{gl(1|2)} = \Pi_{gl(1|2)} \, H_{XX}^{gl(2|2)} \, \Pi_{gl(1|2)}
\label{eq:XXdoll}
\end{equation}

As illustrated in the previous example, the $gl(1|2)$ Hubbard Hamiltonian
is constructed by coupling two copies of the XX Hamiltonian, with fermionic
oscillators $c_{\sigma,i}^\dagger$ and $c_{\sigma,i}$,
$\sigma=\uparrow,\downarrow$. It reads therefore 
\begin{eqnarray}
H_{Hub}^{gl(1|2)} &\!\!=\!\!& \sum_{i=1}^{L} \Big\{ \; \sum_{\sigma=\uparrow,\downarrow}
\Big( c_{\sigma,i}'^\dagger c_{\sigma,i+1}^\dagger c_{\sigma,i}
c_{\sigma,i+1}' + c_{\sigma,i}^\dagger c_{\sigma,i+1}'^\dagger
c_{\sigma,i}' c_{\sigma,i+1} \, - \big( c_{\sigma,i}^\dagger c_{\sigma,i+1}
+ c_{\sigma,i+1}^\dagger c_{\sigma,i} \big) n_{\sigma,i}' n_{\sigma,i+1}'
\Big) \nonumber \\
&& \qquad + \; U (n_{\uparrow,i}'-n_{\uparrow,i}-n_{\uparrow,i}n_{\uparrow,i}')
(n_{\downarrow,i}'-n_{\downarrow,i}-n_{\downarrow,i}n_{\downarrow,i}')
\Big\}
\label{eq:hamHubgl12}
\end{eqnarray}
the space of states at each site $i$ being spanned by tensoring the up
states $|\!\uparrow\rangle_i, |\!\uparrow'\rangle_i,
|\!\uparrow\uparrow'\rangle_i$ with the down states $|\!\downarrow\rangle_i,
|\!\downarrow'\rangle_i, |\!\downarrow\downarrow'\rangle_i$, where
$|\sigma\rangle_i \equiv c_{\sigma,i}^\dagger |0\rangle_i$,
$|\sigma'\rangle_i \equiv c_{\sigma,i}'^\dagger |0\rangle_i$ and
$|\sigma\sigma'\rangle_i \equiv c_{\sigma,i}^\dagger c_{\sigma,i}'^\dagger
|0\rangle_i$ with $\sigma=\uparrow,\downarrow$. It follows that the space of
states of the $gl(1|2)$ Hubbard model is obtained from the space of states
of the $gl(2|2)$ Hubbard model by acting with the projector 
\begin{equation}
\wt\Pi_{gl(1|2)} = \prod_{i=1}^{L}
(n_{\uparrow,i}+n_{\uparrow,i}'-n_{\uparrow,i}n_{\uparrow,i}')
(n_{\downarrow,i}+n_{\downarrow,i}'-n_{\downarrow,i}n_{\downarrow,i}')
\end{equation}
Again, one has
\begin{equation}
H_{Hub}^{gl(1|2)} = \wt\Pi_{gl(1|2)} \, H_{Hub}^{gl(2|2)} \, \wt\Pi_{gl(1|2)}
\end{equation}
which is a direct consequence of (\ref{eq:XXdoll}) and the trivial
embedding of the $gl(1|2)$ and $gl(2|2)$ $C$ matrices entering the
definition of the potential term. \\
Introducing the notation $X_{ij} = \sum_{\sigma=\uparrow,\downarrow}
c_{\sigma,i}^\dagger c_{\sigma,j} \, \cN_{\sigma,ij}'$, see eq.
(\ref{eq:Xij}), where now $\cN_{\sigma,ij}' = c_{\sigma,i}'^\dagger
c_{\sigma,j}' - n_{\sigma,i}'n_{\sigma,j}'$, one has
\begin{equation}
H_{Hub}^{gl(1|2)} = \sum_{i=1}^{L} (X_{i,i+1}+X_{i+1,i}) + U
(n_{\uparrow,i}'-n_{\uparrow,i}-n_{\uparrow,i}n_{\uparrow,i}')
(n_{\downarrow,i}'-n_{\downarrow,i}-n_{\downarrow,i}n_{\downarrow,i}')
\end{equation}
At site $i$, among the nine possible states, four of them have an
interaction energy $-U$ and the other five have an interaction energy $+U$.
These four states are characterised by the constraint $n_{\uparrow,i} +
n_{\downarrow,i} = 1$, hence the projector $\Pi_0$ in the effective
Hamiltonian is again given by (\ref{eq:projP0}) and the relations
(\ref{eq:condP0}) are still satisfied. Therefore the effective Hamiltonian
at second order of perturbation reads 
\begin{eqnarray}
H_{\textrm{eff}}^{(2)} &\!\!=\!\!&-\frac{1}{2U}\sum_{i=1}^{L} \Pi_0
(X_{i,i+1}X_{i+1,i} + X_{i+1,i}X_{i,i+1}) \Pi_0 \nonumber \\
&\!\!=\!\!& - \frac{1}{U}\sum_{i=1}^{L} \Pi_0 \left[ (\half - 2
  S_{i}^z S_{i+1}^z) - (S_{i}^+ S_{i+1}^- + S_{i}^- S_{i+1}^+) \,
  \wt\cN_{\uparrow,i i+1}' \, \wt\cN_{\downarrow,i i+1}' \right] \Pi_0
\, \wt\Pi_{gl(1|2)} \qquad 
\label{effect}
\end{eqnarray}
where the `dressing' factor $\wt\cN_{\sigma,ij}'$ is obtained from
$\cN_{\sigma,ij}'$ by action of the projector $\wt\Pi_{gl(1|2)}$:
$\wt\cN_{\sigma,ij}' = c_{\sigma,i}'^\dagger c_{\sigma,j}' +
c_{\sigma,j}'^\dagger c_{\sigma,i}' - n_{\sigma,i}'n_{\sigma,j}'$.

Unfortunately, the symmetry of this Hamiltonian is not $gl(1|2) \oplus
gl(1|2)$ as one might have hoped from the $gl(2|2)$ case, but only
$gl(1|1) \oplus U(1) \oplus gl(1|1) \oplus U(1)$. The corresponding
bosonic generators are given by $\rho_\sigma^z$, $\eta_\sigma^z$,
$N_\sigma$ (with $\sigma=\uparrow,\downarrow$) and the fermionic ones by
$\phi_{\uparrow\uparrow}^\pm$ and $\phi_{\downarrow\downarrow}^\pm$.

The spectrum of this Hamiltonian is completely contained in the
$gl(2|2)$ case (\ref{eq:hamhubeff}) already described in the previous section.

\subsection{$gl(4|4)$ Hamiltonians}

In the $gl(4|4)$ case, the generators $X_{i;a\beta\gamma,\alpha'\beta'\gamma'}$
at each site $i$ are given by
\begin{equation}
X_{i;\alpha\beta\gamma,\alpha'\beta'\gamma'} =
(-1)^{([\alpha]+[\alpha'])([\beta]+[\gamma])+([\beta]+[\beta'])[\gamma]} \,
X_{i;\alpha\alpha'} \; X'_{i;\beta\beta'} \; X''_{i;\gamma\gamma'}
\end{equation}
The $X_{\alpha\beta\gamma,\alpha'\beta'\gamma'}$ generators are mapped on 
the $E_{pq}$ matrices with the following assignment of indices 
($\alpha,\beta,\gamma,\alpha',\beta',\gamma'=1,2$ and $p,q=1\ldots8$): 
\begin{equation}
111 \to 1, \; 112 \to 6, \; 121 \to 7, \; 122 \to 4, \; 211 \to 5, \; 212
\to 2, \; 221 \to 3, \; 222 \to 8
\end{equation}
Again, this assignment respects the grading in the sense that if
$(\alpha\beta\gamma,\alpha'\beta'\gamma') \to (p,q)$,
$X_{\alpha\beta\gamma,\alpha'\beta'\gamma'}$ and $E_{pq}$ have the same
grade. \\
Then the $gl(4|4)$ XX Hamiltonian (\ref{eq:XXHam}) reads as (the subscripts 
correspond to the site indices):
\begin{eqnarray}
H_{XX}^{gl(4|4)} &\!\!=\!\!& \sum_{i=1}^{L} \big( c_{i}^\dagger c_{i+1} + c_{i+1}^\dagger
c_{i} + 1 - n_{i} - n_{i+1} \big) \Big( c_{i}'^\dagger c_{i+1}'
c_{i}''^\dagger c_{i+1}'' + c_{i+1}'^\dagger c_{i}' c_{i+1}''^\dagger
c_{i}'' \nonumber \\
&& - \, n_{i}' n_{i+1}' (c_{i}''^\dagger c_{i+1}'' + c_{i+1}''^\dagger
c_{i}'') - n_{i}'' n_{i+1}'' (c_{i}'^\dagger c_{i+1}' + c_{i+1}'^\dagger
c_{i}') \Big)
\end{eqnarray}
As in the $gl(2|2)$ case, one gets for the $gl(4|4)$ Hubbard Hamiltonian:
\begin{eqnarray}
H_{Hub}^{gl(4|4)} &\!\!=\!\!& \sum_{i=1}^{L} \; \Big\{ \;
\sum_{\sigma=\uparrow,\downarrow} \big( c_{\sigma,i}^\dagger c_{\sigma,i+1}
+ c_{\sigma,i+1}^\dagger c_{\sigma,i} + 1 - n_{\sigma,i} - n_{\sigma,i+1}
\big) \Big( c_{\sigma,i}'^\dagger c_{\sigma,i+1}' c_{\sigma,i}''^\dagger
c_{\sigma,i+1}'' \nonumber \\
&& + \; c_{\sigma,i+1}'^\dagger c_{\sigma,i}' c_{\sigma,i+1}''^\dagger
c_{\sigma,i}'' - n_{\sigma,i}' n_{\sigma,i+1}' (c_{\sigma,i}''^\dagger
c_{\sigma,i+1}'' + c_{\sigma,i+1}''^\dagger c_{\sigma,i}'') \nonumber \\
&& - \; n_{\sigma,i}'' n_{\sigma,i+1}'' (c_{\sigma,i}'^\dagger
c_{\sigma,i+1}' + c_{\sigma,i+1}'^\dagger c_{\sigma,i}') \Big) + U
(1-2n_{\uparrow,i}'n_{\uparrow,i}'')
(1-2n_{\downarrow,i}'n_{\downarrow,i}'') \; \Big\}
\end{eqnarray}
One observes that this Hamiltonian exhibits a `Russian doll' structure.
Indeed, there are four sectors in the space of states where the $gl(4|4)$
Hamiltonian reduces to the $gl(2|2)$ one. These sectors are defined 
respectively by $n_{\uparrow,i}' = n_{\downarrow,i}' = 1$, $n_{\uparrow,i}'
= n_{\downarrow,i}'' = 1$, $n_{\uparrow,i}'' = n_{\downarrow,i}' = 1$,
$n_{\uparrow,i}'' = n_{\downarrow,i}'' = 1$ for $1 \le i \le L$. The 
obtained Hamiltonian can be further reduced to $gl(1|1)$ Hamiltonian by 
imposing on each site $n_{\sigma,i}=0$ or $n_{\sigma,i}=1$. 

\section{Conclusion and perspectives}
We have constructed super-Hubbard models based on the superalgebras
$gl(N|M)$, with a special focus on models that may apply to SYM theories.
We have seen that in the case of superalgebras, the Jordan--Wigner
transformation is a local isomorphism. Therefore, the interpretation
of the models in terms of `electrons' is more natural.

The symmetry superalgebra and the Hamiltonian have been given, and we
performed a perturbative calculation \emph{\`a la} Klein and Seitz
\cite{klstz} for the Hamiltonians based on the superalgebras $gl(1|2)$ and
$gl(2|2)$. 

The next step in the study of our models is the determination of the spectrum
and the Bethe equations, as they were constructed for Hubbard or
generalisation, using the algebraic Bethe ansatz
\cite{YueDegu,RamMar,martins,Hubsu4}. This is an heavy calculation which we
postpone for further publication, but from the analytical Bethe ansatz
approach, one can guess their form. In particular, as for spin chain
models, one expects as many presentations of the Bethe equations as there
are inequivalent Dynkin diagrams. All these presentations should lead to
the same spectrum. For more informations, we refer to \cite{selene,RS} where
similar calculations were performed in the case of XXX super spin chains.

\section*{Acknowledgments}
GF thanks INFN for a post-doctoral fellowship and for financial support.
ER thanks F. Goehmann for helpful and stimulating discussions, and for
providing useful references.
This work is partially supported by the EC Network
`EUCLID. Integrable models and applications: from strings to condensed
matter', contract number HPRN-CT-2002-00325 and by the ANR project,
`Theories de jauge superconformes', number BLAN06-3 143795.



\end{document}